\documentclass[11pt,english,superscriptaddress,showpacs,aps]{revtex4-1}
\usepackage[latin9]{inputenc}
\usepackage{babel}
\usepackage{amsmath}
\usepackage{amssymb}
\usepackage{esint}
\usepackage[unicode=true,
 bookmarks=true,bookmarksnumbered=false,bookmarksopen=false,
 breaklinks=false,pdfborder={0 0 1},backref=false,colorlinks=false]
 {hyperref}

\makeatletter
\usepackage{graphicx}
\usepackage{color}

\usepackage{slashed}

\usepackage{babel}

\makeatother

\begin{document}

\title{Generation of higher derivatives operators and electromagnetic wave
propagation in a Lorentz-violation scenario}

\author{L. H. C. Borges}

\email{luizhenriqueunifei@yahoo.com.br}

\author{A. G. Dias}

\email{alex.dias@ufabc.edu.br}

\author{A. F. Ferrari}

\email{alysson.ferrari@ufabc.edu.br}

\affiliation{Universidade Federal do ABC, Centro de Ciências Naturais e Humanas,
Av. dos Estados, 5001, Santo André, SP, Brazil, 09210-580. }

\author{J. R. Nascimento}

\email{jroberto@fisica.ufpb.br}

\author{A. Yu. Petrov}

\email{petrov@fisica.ufpb.br}

\affiliation{Departamento de Física, Universidade Federal da Paraíba, Caixa Postal
5008, João Pessoa, Paraíba, Brazil, 58051-970.}
\begin{abstract}
We study the perturbative generation of higher-derivative Lorentz
violating operators as quantum corrections to the photon effective
action, originated from a specific Lorentz violation background, which
has already been studied in connection with the physics of light pseudoscalars.
We calculate the complete one loop effective action of the photon
field through the proper-time method, using the zeta function regularization.
This result can be used as a starting point to study possible effects
of the Lorentz violating background we are considering in photon physics.
As an example, we focus on the lowest order corrections and investigate
whether they could influence the propagation of electromagnetic waves
through the vacuum. We show, however, that no effects of the kind
of Lorentz violation we consider can be detected in such a context,
so that other aspects of photon physics have to be studied.
\end{abstract}
\maketitle

\section{\label{I}Introduction}

Since the years 1990s, a systematic search for Lorentz symmetry breaking
in a wide range of phenomena has been undertaken by both experimental
and theoretical physicists, using as a fundamental tool the Standard
Model Extension (SME) developed by Kostelecky and collaborators\,\cite{KostColl1,SME}.
The idea is to incorporate in the Standard Model (SM) Lagrangian new
terms involving constant background tensors, thus introducing Lorentz
violation (LV) while maintaining fundamental properties such as renormalizability
and gauge invariance. The extension of this idea to include gravity
was first worked out in\,\cite{Kostel}, and it has been recently
extended to include higher derivative operators involving LV\,\cite{kostel-HD1,kostel-HD2}.
More on the status of this subject can be found for example in\,\cite{6thCPT}.

One natural question is the origin of such a multitude of LV terms
added to the SM Lagrangian. They might arise from a spontaneous breaking
of Lorentz symmetry in a more fundamental theory, such as String Theory\,\cite{kosteleckystrings},
but they could also be derived as quantum corrections in an extended
theory of electromagnetic, scalar or gravitational fields involving
couplings of these fields to heavy spinor fields in a Lorentz breaking
manner. This idea was developed in\,\cite{JK}, where one of the
simplest Lorentz-breaking extensions of electrodynamics, the CFJ model\,\cite{CFJ},
was shown to arise in the low-energy effective action of a particular
LV model where a massive fermion is integrated out. This idea was
extended to the perturbative generation of other LV terms: in\,\cite{axion},
for example, we have shown how a model including a massive fermion
that couples in a LV way to the gauge field $A^{\mu}$ and to a pseudoscalar
$\phi$ can generate the coupling between an axion-like particle and
the photon, which is very relevant for contemporary experimental searches
for axions or other light pseudoscalars. Our aim now is to study the
physical consequences of one of the couplings introduced in\,\cite{axion},
this time focusing on the pure Maxwell sector of the theory.

It is known that a very efficient tool for obtaining the complete
low-energy one-loop effective action is the proper-time method\,\cite{Schwinger}
(for a review, see for example\,\cite{BSD}), which was adapted for
gauge theories in\,\cite{mcarthur}. Within the study of the Lorentz
violating theories, the only example we know of the application of
this method is given in\,\cite{ourptime} where it was used to obtain
the non-Abelian CFJ and the gravitational Chern-Simons terms as perturbative
corrections in some LV models. 

In this work, we adopt the version of the proper-time method based
on the zeta function regularization to obtain the complete low-energy
effective action of the gauge field in the Lorentz-breaking extension
of QED which we considered\,\cite{axion}. The effective action thus
calculated can be used as a starting point for the study of further
physical effects of these LV couplings. With this idea in mind, we
look for the influence of the LV on the propagation of electromagnetic
waves in the one-loop corrected theory. It is known that certain Lorentz-breaking
extensions of QED display birefringence and rotation of the polarization
plane of electromagnetic waves in the vacuum\,\cite{waves1,waves2,waves3,waves4,waves5}.
Furthermore, different issues related to wave propagation in various
Lorentz-breaking extensions of QED were studied in a number of papers\,\cite{class1,class2,class3,class4,class5,class6,class7,class8,class9,class10}.
However, there is up to now a very small number of studies of wave
propagation in higher-derivative Lorentz-breaking theories, such as\,\cite{CMR1,CMR2,CMR3},
where Lorentz-breaking modifications up to third order in derivatives
were studied, and issues related to unitarity and causality were considered.
It is worth to mention that other higher-derivative contributions
to the effective action, also up to the third order in derivatives,
were shown to emerge as quantum corrections\,\cite{ourHD1,ourHD2,ourHD3}. 

Considering our model in particular, the one loop corrections we take
into account are the standard nonlinear corrections to the Maxwell
theory, known as the Euler-Heisenberg Lagrangian, with the contributions
of first order in the LV parameters. Since there are active precision
experiments involving photons propagating in a strong magnetic field\,\,\cite{PVLAS1,PVLAS2},
this study could pinpoint a window of opportunity to probe for the
LV background we consider. We write the modified Maxwell's equations
in vacuum, but find out that no modification due to LV in the wave
propagation appears in our model. However, we point out that an experimental
bound on the mass $M$ could in principle be derived once these experiments
are able to measure the nonlinear corrections with sufficient precision.

In section\,\ref{II}, we shall describe the Lorentz-breaking extension
of QED that we consider, which is a subset of the model considered
in\,\cite{axion}. The calculation of the one-loop low-energy effective
action for the photon field is presented in section\,\ref{III}.
The first order corrections generated in the Maxwell action are made
explicit in section\,\ref{IV}, and there we show the LV does not
modify wave propagation in the vacuum. Section\,\ref{V} summarizes
our conclusions and perspectives.

\section{\label{II}The model}

We consider a high energy model containing the photon field $A_{\mu}$
and a single massive charged fermion field $\psi$. The Lagrangian
describing our model is given by 
\begin{equation}
\mathcal{L}=-\ \frac{1}{4}F^{\mu\nu}F_{\mu\nu}+\bar{\psi}\left[i\slashed{\partial}-M-\gamma^{\mu}(gA_{\mu}+F_{\mu\nu}d^{\nu})\right]\psi\,,\label{model}
\end{equation}
where $F_{\mu\nu}=\partial_{\mu}A_{\nu}-\partial_{\nu}A_{\mu}$ is
the field strength, $M$ is the fermion mass and $g$ is the electric
charge. Within our model, LV enters via the nonminimal CPT-breaking
interaction $d^{\nu}\thinspace F_{\mu\nu}\overline{\psi}\gamma^{\mu}\psi$,
where $d^{\mu}$ is a constant background vector (not an axial vector
as in\,\cite{CFJ}). Our main interest in this particular interaction
comes from the fact that it was shown to lead to relevant results
for the search of axion-like particles\,\cite{axion}, which motives
us to investigate further physical consequences of this term. 

Since we assume the fermion mass to be very high, in the low-energy
regime we can integrate in the fermion field, expressing the one loop
effective action $S_{eff}^{(1)}\left[A\right]$ by means of a functional
trace as follows, 
\begin{equation}
S_{eff}^{(1)}\left[A\right]=-\,{\rm Tr}\,\ln\left(i\slashed\partial-M-\gamma^{\mu}\widetilde{A}_{\mu}\right)\,,\label{trace}
\end{equation}
where 
\begin{equation}
\widetilde{A}_{\mu}=gA_{\mu}+F_{\mu\nu}d^{\nu}\ .\label{coupling}
\end{equation}
In terms of the operator $\widetilde{D}_{\mu}=\partial_{\mu}+i\,\widetilde{A}_{\mu}$,
the effective action in Eq.\,(\ref{trace}) can be written as 
\begin{equation}
S_{eff}^{(1)}\left[A\right]=-\,{\rm Tr}\,\ln\left(i\widetilde{\slashed D}-M\right)\,.\label{trace1}
\end{equation}
We note that the proper-time method requires the use of an even-order
differential operator\,\cite{BSD}, so we rewrite the trace on the
previous equation as follows,
\begin{align}
{\rm Tr}\,\ln\left(i\widetilde{\slashed D}-M\right) & =\frac{1}{2}{\rm Tr}\left[\ln\left(i\widetilde{\slashed D}-M\right)+\ln\left(i\widetilde{\slashed D}-M\right)\right]\nonumber \\
 & =\frac{1}{2}{\rm Tr}\left[\ln\left(i\widetilde{\slashed D}-M\right)+\ln\left(-i\widetilde{\slashed D}-M\right)\right]\nonumber \\
 & =\frac{1}{2}{\rm Tr}\,\ln\left(\widetilde{\slashed D}^{2}+M^{2}\right)\,,\label{eq:traceq}
\end{align}
 where in the second line we have inserted $\gamma_{5}\gamma_{5}=I$
and used of the cyclicity of the trace. Thus we can write 
\begin{equation}
S_{eff}^{(1)}\left[A\right]=-\frac{1}{2}\,\mathrm{Tr\,}\ln\left(\widetilde{D}^{2}+i\,\Sigma^{\mu\nu}\widetilde{F}_{\mu\nu}+M^{2}\right)\,,\label{trace2}
\end{equation}
where 
\begin{equation}
\Sigma^{\mu\nu}=\frac{1}{4}\left[\gamma^{\mu},\gamma^{\nu}\right]\:;\:\widetilde{F}_{\mu\nu}=\partial_{\mu}\widetilde{A}_{\nu}-\partial_{\nu}\widetilde{A}_{\mu}\,.\label{defi}
\end{equation}
This form of the one loop effective action is suitable for the evaluation
by means of the proper-time method, as we show in the next section.

\section{\label{III}Evaluation of Quantum corrections}

We calculate now the explicit form of the one-loop contribution to
the effective action in Eq.\,\eqref{trace2} using the zeta function
regularization (for a review on this approach, see\,\cite{zeta}).
The zeta function $\zeta(s)$ is an important tool used to represent
functional determinants in quantum field theory, and it is written
in terms of the integral over proper time \textbf{$\tau$} as follows
\begin{equation}
\zeta\left(s\right)=\frac{1}{\Gamma\left(s\right)}\int_{0}^{\infty}\tau^{s-1}\mathrm{Tr}\left(e^{-\mathcal{O}\tau}\right)d\tau\,,\label{zeta}
\end{equation}
where $\mathcal{O}$ represents an dimensionless differential operator.
It follows from this expression that $\left(\frac{d\zeta}{ds}\right)\Bigg|_{s=0}=\mathrm{Tr}\left(\ln\mathcal{O}\right),$
and therefore 
\begin{equation}
S_{eff}^{(1)}\left[A\right]=-\frac{1}{2}\,\left(\frac{d\zeta}{ds}\right)\Bigg|_{s=0}\ ,\label{zeta2}
\end{equation}
where, for the sake of our work, 
\begin{equation}
\mathcal{O}\sim\widetilde{D}^{2}+i\,\Sigma^{\mu\nu}\widetilde{F}_{\mu\nu}+M^{2}\ .\label{operator}
\end{equation}
Equation\,(\ref{zeta2}) summarizes the use of the zeta function
method for obtaining the one-loop quantum corrections. Up to our knowledge,
although it has been applied in many physically interesting models
including gravity\,\cite{zeta} and supersymmetric field theories\,\cite{mcarthur},
this method was never used for the study of Lorentz-breaking theories.

Now, let us briefly describe how we apply the zeta function approach
in our model, following\,\cite{mcarthur}. Substituting Eq.\,(\ref{operator})
in Eq.\,(\ref{zeta}), we have 
\begin{equation}
\zeta\left(s\right)=\frac{1}{\Gamma\left(s\right)}\int_{0}^{\infty}d\tau\,\tau^{s-1}e^{-\frac{m^{2}\tau}{\mu^{2}}}\,\widetilde{K}\left(\frac{\tau}{\mu^{2}}\right)\,,\label{zeta3}
\end{equation}
where we identify the kernel trace 
\begin{equation}
\widetilde{K}\left(\frac{\tau}{\mu^{2}}\right)=e^{-\frac{\tau}{\mu^{2}}\widetilde{D}^{2}}\,\mathrm{tr}\left\{ e^{-i\,\frac{\tau}{\mu^{2}}\Sigma^{\mu\nu}\widetilde{F}_{\mu\nu}}\right\} \,.\label{kernel}
\end{equation}
Here the trace is taken over spin indices, and the $\mu$ parameter
has mass dimension one and is introduced to make the differential
operator in Eq.\,(\ref{operator}) dimensionless. The kernel trace
can be related with the local kernel through the expression 
\begin{equation}
\widetilde{K}\left(\tau\right)=\int d^{4}x\,\,\mathrm{tr}\left\{ e^{-i\,\tau\Sigma^{\mu\nu}\widetilde{F}_{\mu\nu}}\right\} \,K\left(\tau\right)\,,\label{kernel1}
\end{equation}
with 
\begin{equation}
K\left(\tau\right)=\lim_{x\rightarrow x'}e^{-\tau\widetilde{D}^{2}}\delta^{4}\left(x-x'\right)\,.\label{kernel2}
\end{equation}
Substituting Eq.\,(\ref{kernel1}) in Eq.\,(\ref{zeta3}), we obtain
\begin{equation}
\zeta\left(s\right)=\frac{1}{\Gamma\left(s\right)}\int_{0}^{\infty}d\tau\,\tau^{s-1}e^{-\frac{m^{2}\tau}{\mu^{2}}}\int d^{4}x\,\mathrm{tr}\left\{ e^{-i\,\frac{\tau}{\mu^{2}}\Sigma^{\mu\nu}\widetilde{F}_{\mu\nu}}\right\} \,K\left(\frac{\tau}{\mu^{2}}\right)\,.\label{zeta4}
\end{equation}
To find an explicit expression for $K\left(\tau\right)$, one may
repeat all the steps described in\,\cite{mcarthur}, arriving at
\begin{equation}
K\left(\tau\right)=\frac{1}{16\pi^{2}\tau^{2}}\,\det\left[\frac{-\tau i\widetilde{F}}{\sinh\left(-\tau i\widetilde{F}\right)}\right]^{1/2}\,,\label{det}
\end{equation}
where in this expression the tensor $\tilde{F}$ is considered as
a four by four matrix in calculating the determinant. Therefore, 
\begin{equation}
\zeta\left(s\right)=\frac{\mu^{4}}{16\pi^{2}}\,\frac{1}{\Gamma\left(s\right)}\int_{0}^{\infty}d\tau\,\tau^{s-3}e^{-\frac{m^{2}\tau}{\mu^{2}}}\int d^{4}x\,\mathrm{tr}\left\{ e^{-i\,\frac{\tau}{\mu^{2}}\Sigma^{\mu\nu}\widetilde{F}_{\mu\nu}}\right\} \,\det\left[\frac{-\tau i\widetilde{F}}{\mu^{2}\sinh\left(-\frac{\tau}{\mu^{2}}i\widetilde{F}\right)}\right]^{1/2}\,.\label{zeta5}
\end{equation}
To calculate the trace and the determinant in the right-hand side
of the Eq.\,(\ref{zeta5}), we carry out the expansions 
\begin{equation}
\det\left[\frac{-\tau i\widetilde{F}}{\mu^{2}\sinh\left(-\frac{\tau}{\mu^{2}}i\widetilde{F}\right)}\right]^{1/2}=1-\frac{\tau^{2}}{12\mu^{4}}\widetilde{F}^{2}+\frac{\tau^{4}}{144\mu^{8}}\left[\frac{2}{5}\mathrm{Tr}\left(\widetilde{F}^{4}\right)+\frac{1}{2}\widetilde{F}^{4}\right]+\cdots\ ,
\end{equation}
where $\mbox{Tr}$ means a trace over the spacetime indices of $\tilde{F}$,
i.e., 
\begin{equation}
\mathrm{Tr}\left(\widetilde{F}^{2}\right)=\widetilde{F}^{\mu\nu}\widetilde{F}_{\nu\mu}=-\widetilde{F}^{2}\,,\quad\mathrm{Tr}\left(\widetilde{F}^{4}\right)=\widetilde{F}^{\mu\nu}\widetilde{F}_{\nu\alpha}\widetilde{F}^{\alpha\beta}\widetilde{F}_{\beta\mu}\,,
\end{equation}
and
\begin{align}
\mathrm{tr}\left\{ e^{-i\,\frac{\tau}{\mu^{2}}\Sigma^{\mu\nu}\widetilde{F}_{\mu\nu}}\right\} = & \mathrm{tr}\left(I\right)-i\,\frac{\tau}{\mu^{2}}\mathrm{tr}\left(\Sigma^{\mu\nu}\widetilde{F}_{\mu\nu}\right)-\frac{\tau^{2}}{2\mu^{4}}\mathrm{tr}\left(\Sigma^{\mu\nu}\widetilde{F}_{\mu\nu}\right)^{2}\nonumber \\
 & +i\,\frac{\tau^{3}}{6\mu^{6}}\mathrm{tr}\left(\Sigma^{\mu\nu}\widetilde{F}_{\mu\nu}\right)^{3}+\frac{\tau^{4}}{24\mu^{8}}\mathrm{tr}\left(\Sigma^{\mu\nu}\widetilde{F}_{\mu\nu}\right)^{4}+\cdots\,.
\end{align}
It can be shown that
\begin{equation}
\mathrm{tr}\left(\Sigma^{\mu\nu}\widetilde{F}_{\mu\nu}\right)^{2}=-2\widetilde{F}^{2}\,;\;\mathrm{tr}\left(\Sigma^{\mu\nu}\widetilde{F}_{\mu\nu}\right)^{4}=\widetilde{F}^{4}-\left(^{*}\widetilde{F}^{\mu\nu}\widetilde{F}_{\mu\nu}\right)^{2}\,,\label{eq:tracespinor1}
\end{equation}
whereas traces of an odd number of $\Sigma$ matrices vanish, and
the dual electromagnetic tensor is defined as $^{*}\widetilde{F}^{\mu\nu}=\frac{1}{2}\epsilon^{\mu\nu\alpha\beta}\widetilde{F}_{\alpha\beta}$.
Putting all this together, we obtain the one-loop quantum corrections
to the effective action of the photon field $A_{\mu}$ in the form
\begin{multline}
S_{eff}^{(1)}\left[A\right]=-\frac{1}{48\pi^{2}}\int d^{4}x\,\Biggl\{-\ln\left(\frac{M^{2}}{\mu^{2}}\right)\widetilde{F}^{2}+\frac{1}{8M^{4}}\Biggl[\frac{2}{15}\mathrm{Tr}\left(\widetilde{F}^{4}\right)-\frac{1}{3}\widetilde{F}^{4}\\
-\frac{1}{2}\left(^{*}\widetilde{F}^{\mu\nu}\widetilde{F}_{\mu\nu}\right)^{2}\Biggr]+\frac{1}{16M^{8}}\widetilde{F}^{2}\left[\frac{2}{5}\mathrm{Tr}\left(\widetilde{F}^{4}\right)+\frac{1}{2}\left(^{*}\widetilde{F}^{\mu\nu}\widetilde{F}_{\mu\nu}\right)^{2}\right]\\
+\frac{5}{96M^{12}}\left[\widetilde{F}^{4}-\left(^{*}\widetilde{F}^{\mu\nu}\widetilde{F}_{\mu\nu}\right)^{2}\right]\left[\frac{2}{5}\mathrm{Tr}\left(\widetilde{F}^{4}\right)+\frac{1}{2}\widetilde{F}^{4}\right]+\cdots\cdots\Biggr\}\,.\label{eq:seffpronta}
\end{multline}
The Lorentz violation is implicitly taken into account inside the
${\widetilde{F}}_{\mu\nu}$, since from\,(\ref{coupling}) and\,(\ref{defi}),
we have 
\begin{equation}
\widetilde{F}_{\mu\nu}=gF_{\mu\nu}+d^{\lambda}\left(\partial_{\mu}F_{\nu\lambda}-\partial_{\nu}F_{\mu\lambda}\right)\ ,\label{LV}
\end{equation}
and, for example
\begin{align}
\widetilde{F}^{2}=\widetilde{F}_{\mu\nu}\widetilde{F}^{\mu\nu} & =g^{2}F_{\mu\nu}F^{\mu\nu}+4gd_{\alpha}F_{\mu\nu}\left(\partial^{\mu}F^{\nu\alpha}\right)\nonumber \\
 & +2d^{\lambda}d_{\alpha}\left[\left(\partial_{\mu}F_{\nu\lambda}\right)\left(\partial^{\mu}F^{\nu\alpha}\right)-\left(\partial_{\mu}F_{\nu\lambda}\right)\left(\partial^{\nu}F^{\mu\alpha}\right)\right]\,.\label{FF}
\end{align}
The Lorentz violation causes a space-time anisotropy that leads us
to a preferred frame, in other words, we expect a dependence of physical
measurements on the directions of the Lorentz violation parameters.
To make evident such dependence, one may rewrite Eq.\,(\ref{FF})
explicitly in terms of the electric and magnetic fields ${\bf E}$
and ${\bf B}$ respectively, up to the first order in the LV parameter,
as follows 
\begin{multline}
\widetilde{F}^{2}=2g^{2}\left({\bf B}^{2}-{\bf E}^{2}\right)+4g\left\{ \left(d^{0}\right)\left[{\bf B}\cdot\left({\bf \nabla\times}{\bf {\bf E}}\right)+{\bf E}\cdot\left(\partial_{0}{\bf E}\right)\right]\right.\\
-{\bf d}\cdot\left[{\bf E}\times\left(\partial_{0}{\bf B}\right)\right]+{\bf E}\cdot\left[{\bf \nabla}\left({\bf d}\cdot{\bf E}\right)\right]+\left({\bf \nabla}\cdot{\bf B}\right)\left({\bf d}\cdot{\bf B}\right)\\
\left.-{\bf B}\cdot\left[{\bf \nabla}\left({\bf d}\cdot{\bf B}\right)\right]+{\bf d}\cdot\left[\left({\bf \nabla}\times{\bf B}\right)\times{\bf B}\right]\right\} 
\end{multline}
thus unveiling the dependence of the action on the orientation of
the the fields ${\bf E}$ and ${\bf B}$ with respect to the Lorentz
violation parameter $d^{\nu}=\left(d^{0},\ {\bf d}\right)$.

Summing up the results of this section, the full effective action
$S_{eff}\left[A\right]$ can be represented in the following way 
\begin{alignat}{1}
S_{eff}\left[A\right]=\int d^{4}x\Biggl[ & -\frac{1}{4}F^{\mu\nu}F_{\mu\nu}+\frac{g^{2}}{48\pi^{2}}\ln\left(\frac{M^{2}}{\mu^{2}}\right)F^{\mu\nu}F_{\mu\nu}+{\cal L}_{F^{4}}\nonumber \\
 & +\frac{g}{12\pi^{2}}\ln\left(\frac{M^{2}}{\mu^{2}}\right)d_{\alpha}F_{\mu\nu}\left(\partial^{\mu}F^{\nu\alpha}\right)+\cdots\Biggr]\ .\label{eq:seffA}
\end{alignat}
The second term in\,\eqref{eq:seffA} is similar to the standard
vacuum polarization correction to the Maxwell theory. The third is
the lowest order non-linear correction to the Maxwell theory, which
in the case of QED is known in literature by the name of Euler-Heisenberg
term (for a review, see f.e.\,\cite{Dunne}). More explicitly, from
Eq.\,\eqref{eq:seffpronta}, we obtain
\begin{equation}
{\cal L}_{F^{4}}=-\frac{1}{8\pi}\cdot\frac{1}{48\pi M^{4}}\left[\frac{2}{15}\mathrm{Tr}\left(\widetilde{F}^{4}\right)-\frac{1}{3}\widetilde{F}^{4}-\frac{1}{2}\left(^{*}\widetilde{F}^{\mu\nu}\widetilde{F}_{\mu\nu}\right)^{2}\right]\Bigg|_{d^{\nu}=0}\thinspace,
\end{equation}
which, by using that\begin{subequations}
\begin{align}
\mathrm{Tr}\left(\widetilde{F}^{4}\right)\Bigg|_{d^{\nu}=0} & =g^{4}\left[2\left({\bf E}^{2}-{\bf B}^{2}\right)^{2}+4\left({\bf E}\cdot{\bf B}\right)^{2}\right]\,,\\
\widetilde{F}^{4}\Bigg|_{d^{\nu}=0} & =4g^{4}\left({\bf E}^{2}-{\bf B}^{2}\right)^{2}\,,\,\left(^{*}\widetilde{F}^{\mu\nu}\widetilde{F}_{\mu\nu}\right)^{2}\Bigg|_{d^{\nu}=0}=16g^{4}\left({\bf E}\cdot{\bf B}\right)^{2}\,,
\end{align}
\end{subequations}can be cast as,
\begin{equation}
{\cal L}_{F^{4}}=\frac{R}{8\pi}\left({\bf E}^{2}-{\bf B}^{2}\right)^{2}+\frac{S}{8\pi}\left({\bf E}\cdot{\bf B}\right)^{2}\thinspace,\label{eq:LEH}
\end{equation}
where 
\begin{equation}
R=\frac{g^{4}}{45\pi M^{4}},\thinspace S=7R\thinspace.\label{eq:RS}
\end{equation}
This reproduces the well known Euler-Heisenberg Lagrangian, the first
nonlinear corrections due to QED, if the charge $g$ and the mass
$M$ are taken as the electron charge $e$ and mass $m$, respectively\,\cite{f3wp}.
The fourth term in\,\eqref{eq:seffA} represents the correction due
to LV in the model. The ellipsis in Eq.\,\eqref{eq:seffA} denotes
higher orders corrections both in the power of fields, and of the
LV parameters, which will not be considered in this work. 

The standard Euler-Heisenberg corrections of QED are quite small effects,
which are nevertheless expected to be probed in the near future, investigating
modifications in the propagation of photons in a region with a strong
magnetic field, as the vacuum magnetic birefringence\,\cite{PVLAS1,PVLAS2}.
The nonlinear effects in our model are actually smaller, since they
involve the mass $M$ which is much larger than the electron mass.
On general, LV effects are also expected to be very tiny, since no
evidence for them has been observed so far. We assume that, in principle,
nonlinear and LV effects in our model are comparable, so that none
will be discarded, and we perform a more robust analysis of the possible
effects of the LV background we are considering. Therefore, we will
not study wave propagation considering only the terms quadratic in
$F$ appearing in Eq.\,\eqref{eq:seffA}, which would amount to study
modifications of wave propagation originated only from the LV, but
we will include also the effects of the non-linearities in the fields,
thus including all the terms presented in Eq.\,\eqref{eq:seffA}
in our analysis.

\section{\label{IV}Electromagnetic wave propagation}

In this section we investigate the possible influence of the Lorentz-violation
background described by Eq.\,\eqref{model} in the electromagnetic
wave propagation in the vacuum. It is convenient to rewrite Eq.\,\eqref{eq:seffA}
as follows, 
\begin{alignat}{1}
S_{eff}\left[A\right]=\int d^{4}x & \left[-\frac{1}{16\pi}\left(1+Cg^{2}\right)F^{\mu\nu}F_{\mu\nu}+{\cal L}_{F^{2}}-\frac{Cg}{4\pi}d_{\alpha}F_{\mu\nu}\left(\partial^{\mu}F^{\nu\alpha}\right)+\cdots\right]\thinspace,\label{eq:Seff2}
\end{alignat}
where $C=-\frac{1}{3\pi}\ln\left(\frac{M^{2}}{\mu^{2}}\right)$. In
this equation, we have changed the normalization of the fields to
coincide with the ones used in\,\cite{f3wp}. The covariant field
equations derived from Eq.\,\ref{eq:Seff2} read
\begin{multline}
\partial_{\mu}\left[\left(1+Cg^{2}\right)F^{\mu\nu}-R\left(F^{\alpha\beta}F_{\alpha\beta}\right)F^{\mu\nu}-\frac{S}{4}\left(^{*}F^{\alpha\beta}F_{\alpha\beta}\right){}^{*}F^{\mu\nu}\right.\\
\left.\vphantom{\frac{S}{4g^{2}}\left(^{*}\widetilde{F}^{\alpha\beta}\widetilde{F}_{\alpha\beta}\right)}+Cgd_{\alpha}\left(\partial^{\alpha}F^{\mu\nu}+\partial^{\mu}F^{\nu\alpha}-\partial^{\nu}F^{\mu\alpha}\right)\right]=0\,.\label{eq:mequation-1}
\end{multline}

The modified Maxwell's equations derived from this action are more
conveniently written in terms of the vectors ${\bf D}$ and ${\bf H}$
defined as\begin{subequations}
\begin{align}
D_{i} & =4\pi\frac{\partial}{\partial E_{i}}\left[-\frac{1}{16\pi}\left(1+Cg^{2}\right)F^{\mu\nu}F_{\mu\nu}+{\cal L}_{F^{4}}\right]\thinspace,\\
H_{i} & =-4\pi\frac{\partial}{\partial B_{i}}\left[-\frac{1}{16\pi}\left(1+Cg^{2}\right)F^{\mu\nu}F_{\mu\nu}+{\cal L}_{F^{4}}\right]\thinspace,
\end{align}
\end{subequations}or, more explicitly, as,\begin{subequations}\label{DH1}
\begin{align}
{\bf {D}} & =\left(1+Cg^{2}\right){\bf {E}}+2R\left({\bf {E}}^{2}-{\bf {B}}^{2}\right){\bf {E}}+S\left({\bf {E}\cdot{\bf {B}}}\right){\bf {B}}\thinspace,\\
{\bf {H}} & =\left(1+Cg^{2}\right){\bf {B}}+2R\left({\bf {E}}^{2}-{\bf {B}}^{2}\right){\bf {B}}+S\left({\bf {E}\cdot{\bf {B}}}\right){\bf {E}\ .}
\end{align}
\end{subequations}With these definitions, the equations of motion
derived from Eq.\,\ref{eq:Seff2} are given by
\begin{equation}
{\bf \nabla}\cdot{\bf D}=-Cg\mathbf{d}\cdot\left\{ -\nabla\times\left(\nabla\times\mathbf{E}+\partial_{0}\mathbf{B}\right)\right\} \thinspace,\label{eq:ME1}
\end{equation}
and
\begin{align}
-\partial_{0}{\bf D}+({\bf \nabla}\times{\bf H}) & =-d^{0}\nabla\times\left[\nabla\times\mathbf{E}+\partial_{0}\mathbf{B}\right]\nonumber \\
 & +\mathbf{d}\cdot\nabla\left[\partial_{0}\mathbf{E}-\nabla\times\mathbf{B}\right]-\mathbf{d}\times\left[\partial_{0}^{2}\mathbf{B}-\nabla^{2}\mathbf{B}\right]\nonumber \\
 & -\nabla\left(\mathbf{d}\cdot\left[\partial_{0}\mathbf{E}-\nabla\times\mathbf{B}\right]\right)\thinspace,\label{eq:ME2}
\end{align}
together with the unmodified equations 
\begin{equation}
{\bf \nabla}\cdot{\bf B}=0\,,\;{\bf \nabla}\times{\bf E}=-\partial_{0}{\bf B}\,.\label{eq:ME3}
\end{equation}

The nonlinearity in the electromagnetic fields is encoded in the left
hand side of Eqs.\,\eqref{eq:ME1} and\,\eqref{eq:ME2}. Due to
the nonlinearity, one cannot simply state that the right hand side
of these equations vanishes in the vacuum, as they would in the linear
electrodynamics. It is true, however, that the nonlinear corrections
to the usual Maxwell equations are of first order in the small coefficients
$R$ and $S$, and since the right hand side of Eqs.\,\eqref{eq:ME1}
and\,\eqref{eq:ME2} is already of first order in the LV parameter
$d^{\mu}$, this nonlinear corrections would amount to a second order
effect, that can be disregarded. The outcome is that the LV parameter
disappears from \eqref{eq:ME1} and\,\eqref{eq:ME2}, which reduce
to\begin{subequations}\label{eq:ME1-1}
\begin{align}
{\bf \nabla}\cdot{\bf D} & =0\thinspace,\\
-\partial_{0}{\bf D}+({\bf \nabla}\times{\bf H}) & =0\thinspace.
\end{align}
\end{subequations}Now these equations are identical in form to the
ones used to find the usual Euler-Heisenberg modifications to wave
propagation (see for example~\cite{Greiner}), so they will predict
the same kind of effects, except in our case we expected them to be
smaller than in the usual QED, since they involve inverse powers of
the large mass $M$. 

As an example, one may consider weak electromagnetic fields ${\bf E}_{P}$
and ${\bf B}_{P}$ propagating in the presence of a constant, strong
magnetic field ${\bf B}_{0}$. More specifically, one assumes
\begin{equation}
{\bf {E}}={\bf {E}}_{P}\ ;{\bf {B}}={\bf {B}}_{0}+{\bf {B}}_{P}\ ,\label{SWP}
\end{equation}
with the conditions
\begin{align}
\vert{\bf {E}}_{P}\vert & \ll(R)^{-1/2},(S)^{-1/2}\ ,\\
\vert{\bf {B}}_{P}\vert & \ll{\bf {B}}_{0}\ll(R)^{-1/2},(S)^{-1/2}\ ,
\end{align}
which allows one to linearize the equations, thus finding plane waves
solutions. In this case, one finds different dispersion relations
depending whether the electric field ${\bf {E}}_{P}$ is perpendicular
or parallel to the constant magnetic field ${\bf {B}}_{0}$,
\begin{equation}
\omega_{\perp}=k\left(1-\frac{7R{B}_{0}^{2}}{2}\right)\ ,\ \ \omega_{\parallel}=k\left(1-2R{B}_{0}^{2}\right)\ ,\label{DYE152}
\end{equation}
that is to say, one finds birefringence in wave propagation. These
results are of the same form as in the usual QED, except they are
much smaller since Eq.\,\ref{eq:RS} involves the fermion mass $M$,
much larger than the electron mass. 

In conclusion, even if wave propagation in vacuum cannot yield any
experimental constrains on the LV parameter $d^{\mu}$, the model
we consider in this work, involving a very massive fermion field which
intermediates the LV appearing in very high energy to the low energy
effective action of the photon field, still generates a correction
in the standard Euler-Heisenberg Lagrangian. Therefore, in principle
a bound on the mass $M$ could be inferred from experimental studies,
insofar these become able to measure the effects described in\,\eqref{DYE152}
with sufficient precision.

\section{\label{V}Conclusions }

The extension of the minimal SME to include higher derivate Lorentz
violating couplings is a new undertaking in a very active area in
theoretical and experimental physics. Even if, by dimensional reasons,
the nonminimal couplings are expected to be smaller than the minimal
ones, they might be relevant in new physical contexts yet under experimental
exploration, such as the interaction between photons and light pseudoscalars\,\cite{axion}.
In this work, we paved the way for further explorations of the physical
consequences of one of the nonminimal couplings discussed in\,\cite{axion},
by calculating the corrections generated by it in the low energy effective
action of the Maxwell field. We used the proper-time method, together
with the zeta function regularization, to integrate out the heavy
fermion responsible for the introduction of LV in our model. With
this result at hand, one may study the consequences of this class
of LV in photon phenomenology. This is an approach that has shown
itself to be rather fertile in the context of the minimal SME, providing
strong phenomenological bounds on several of the SME coefficients.
In our case, we were led to consider the case of photons propagating
in a region with a strong, constant magnetic field, investigating
whether LV effects could compete with the effects of nonlinear effects
induced by the quantum corrections, as it was done in\,\cite{f3wp}
for a specific minimal LV coupling. The end result was that no effects
of the Lorentz violating parameter $d^{\mu}$ can be detected in wave
propagation in vacuum, and the only remnant of the high energy LV
background we considered is actually the correction due to the presence
of the very massive fermion in the Euler-Heisenberg Lagrangian. This
allows in principle to find bounds on the mass of this fermion, assuming
these effects can be measured with sufficient precision. 

These results leave open the question of whether other physical effects
that could be drawn from the modified gauge action we derived in this
work could leave to observable effects, thus providing bounds on these
nonminimal LV couplings. This is not a trivial question since one
should not disregard the standard nonlinear effects that are also
induced from the integration of the massive fermion, and we believe
it is an interesting line to pursue in future works.

\textbf{\medskip{}
}

\textbf{Acknowledgements.} The authors would like to thank the referee
for pointing out a simpler way to derive some of the results presented
in section \eqref{IV}\textbf{.} This work was partially supported
by Conselho Nacional de Desenvolvimento Científico e Tecnológico (CNPq)
and Fundação de Amparo a Pesquisa do Estado de São Paulo (FAPESP),
via the following grants: CNPq 303094/2013-3 (AGD), 303783/2015-0
(AYP) and 482874/2013-9 (AFF); FAPESP 2013/22079-8 (AGD and AFF),
2014/24672-0 (AFF), and 2013/01231-6 (LHCB).

\end{document}